\documentclass[]{elsarticle}

\usepackage{graphicx}
\usepackage{caption,subcaption}

\newcommand{\MT}[2]{\ensuremath{M_{#1}\left[#2\right]}}
\newcommand{\MI}[2]{\ensuremath{M_{#1}^{-1}\left[#2\right]}}

\newcommand{\regd}[1]{\ensuremath{\hat{\partial}_{#1}}}

\newcommand{\HPL}[2]{{\rm H}_{#1}(#2)}
\newcommand{\HS}[2]{{\rm S}_{#1}(#2)}

\usepackage[T1]{fontenc}  
\usepackage{amsmath}
\usepackage{hyperref}

\renewcommand\H{\operatorname{H}}

\renewcommand\S{\operatorname{S}}

\newcommand\ii{\mathbbmsl{i}}
\DeclareMathAlphabet{\mathbbmsl}{U}{bbm}{m}{sl}

\newcommand\prog[1]{{\tt #1}}

\newcommand\cmd[1]{{\tt #1}}

\newenvironment{mmacode}{%
 \small
  \newcommand\IIn[1]{{\scriptsize \tt In[##1]}&{\scriptsize \tt :=} &}%
  \newcommand\OOut[1]{{\scriptsize \tt Out[##1]}&{\scriptsize \tt =} &}%
  \newcommand\In{\stepcounter{mmano}\IIn{\themmano}}%
  \newcommand\Out{\OOut{\themmano}}%
  \newcommand\codewidth{125mm}%
  \begin{center}%
  \begin{tabular}{|@{\hspace{2ex}}rr@{\hspace{2ex}}p{\codewidth}@{\hspace{2ex}}|}%
  \hline
}{%
  \\[-4ex] && \\ \hline
  \end{tabular}%
  \end{center}%
}
\newcounter{mmano}

\begin{document}

\frontmatter

\title{\vskip-3cm{\baselineskip14pt
    \begin{flushleft}
      \normalsize SFB/CPP-13-50\\
      \normalsize TTP13-27\\
      \normalsize LPN13-052
  \end{flushleft}}
  \vskip1.5cm
  \prog{MT}: a \prog{Mathematica} package to compute convolutions
}

\author[kit]{Maik H\"oschele}
\ead{maik.hoeschele@kit.edu}

\author[kit]{Jens Hoff}
\ead{jens.hoff@kit.edu}

\author[kit]{Alexey Pak}
\ead{alexey.pak@iosb.fraunhofer.de}

\author[kit]{Matthias Steinhauser\corref{cor}}
\ead{matthias.steinhauser@kit.edu}

\author[kit]{Takahiro Ueda}
\ead{takahiro.ueda@kit.edu}

\cortext[cor]{Corresponding author}

\address[kit]{Institut f{\"u}r Theoretische Teilchenphysik, Karlsruhe Institute of Technology (KIT), 76128 Karlsruhe, Germany
}

\date{}

\begin{abstract}
  We introduce the \prog{Mathematica} package \prog{MT} which can be
  used to compute, both analytically and numerically, convolutions
  involving harmonic polylogarithms, polynomials or
  generalized functions. As applications contributions to
  next-to-next-to-next-to leading order Higgs boson production and the
  Drell-Yan process are discussed.
\end{abstract}
\maketitle


\newpage


\section*{Program summary}

\begin{itemize}

\item[]{\it Title of program:}
  {\tt MT}

\item[]{\it Available from:}\\
  {\tt
  http://www-ttp.physik.uni-karlsruhe.de/Progdata/ttp13/ttp13-27/
  }

\item[]{\it Computer for which the program is designed and others on which it
    is operable:}
  Any computer where \prog{Mathematica} version~6 or higher is running.

\item[]{\it Operating system or monitor under which the program has been
    tested:}
  Linux

\item[]{\it No. of bytes in distributed program including test data etc.:}
  approximately $50\,000$ bytes, and tables of approximately 60 megabytes

\item[]{\it Distribution format:}
  source code

\item[]{\it Keywords:}
  Convolution of partonic cross sections and splitting functions,
  Mellin transformation, harmonic sums, harmonic polylogarithms,
  Higgs boson production, Drell-Yan process

\item[]{\it Nature of physical problem:}
  For the treatment of collinear divergences connected to initial-state
  radiation it is necessary to consider convolutions of partonic cross
  sections with splitting functions. \prog{MT} can be used to compute such
  convolutions.

\item[]{\it Method of solution:}
  \prog{MT} is implemented in \prog{Mathematica}
  and we provide several functions in order to perform
  transformations to Mellin space, manipulations of the
  expressions, and inverse Mellin transformations.

\item[]{\it Restrictions on the complexity of the problem:}
  In case the weight of the input quantities is too high
  the tables for the (inverse) Mellin transforms have to be extended.
  In the current implementation the tables contain expressions up to weight
  eight, code for the generation of tables of even higher
    weight is provided, too.\\
  \prog{MT} can only handle convolutions of expressions involving
  harmonic polylogarithms, plus distributions and polynomials in the
  partonic variable $x$.

\item[]{\it Typical running time:}
  In general the run time for the individual operations is at most of the
  order of a few minutes (depending on the speed and memory of the computer).

\end{itemize}



\section{Introduction}

The calculation of higher order perturbative corrections within QCD
inevitably contains contributions where quarks and/or gluons are
radiated off initial-state partons. Such processes are
accompanied by infra-red singularities which arise from collinear
emissions.

These divergences must cancel against convolutions of splitting
functions with lower-order cross sections which provide subtraction
terms leading to finite partonic cross sections. The origin of the
subtraction terms are ultra-violet divergences present in the bare
parton densities which, due to the masslessness of all involved partons,
can be viewed as counterterms for collinear divergences in the partonic
cross section~\cite{Ellis:1991qj}.  Alternatively, one can consider the removal
of the divergences as a redefinition of the parton distribution
functions (PDFs).  Since the latter are in general given in the
$\overline{\rm MS}$ scheme we adopt this scheme also for our
calculations.

In this paper we provide a description of the \prog{Mathematica} package
\prog{MT} which can be used to perform convolutions of splitting
functions with partonic cross sections expressed in terms of harmonic
polylogarithms (HPLs), delta functions, plus distributions, polynomials
in $x$, and factors $1/x$, $1/(1-x)$, and $1/(1+x)$. Here $x$ is the
partonic variable typically defined as $x=M^2/s$ where $M$ is the mass
of the (intermediate) final state and $\sqrt{s}$ is the partonic
center-of-mass energy.

The method to compute convolutions implemented in \prog{MT} relies heavily on
the Mellin transformation and its properties.  For a comprehensive discussion
of the Mellin transform and the list of all Mellin images appearing in the
calculation of the next-to-next-to-leading order (NNLO) Higgs boson production
rate see Refs.~\cite{Blumlein:1998if,Ablinger:2011te}.  In contrast, we
decided to relate all required results to a limited set of Mellin transforms
of HPLs with a certain maximum weight.

As applications we discuss convolution contributions to
next-to-next-to-next-to-leading order (N$^3$LO) Higgs boson production
and the Drell-Yan process. In particular we provide the partonic cross
sections up to NNLO expanded sufficiently deep in $\epsilon=(4-D)/2$, 
where $D$ is the number of space-time dimensions. As far as Higgs
boson production is concerned the results have been obtained in
Refs.~\cite{Hoschele:2012xc,Buehler:2013fha}.  The corresponding
results for the Drell-Yan process are new. It is convenient to
introduce the following notation for the partonic cross sections
\begin{eqnarray}
  \tilde\sigma_{ij}(x) &=& A
  \Bigg[
  \tilde\sigma^{(0)}_{ij}(x)
  + \frac{\alpha_s}{\pi} \tilde\sigma^{(1)}_{ij}(x)
  + \left(\frac{\alpha_s}{\pi}\right)^2 \tilde\sigma^{(2)}_{ij}(x)
  + \ldots
  \Bigg]\,,
  \label{eq::sigtil}
\end{eqnarray}
where $A$ collects all constants such that
\begin{eqnarray}
  \tilde\sigma^{(0),{\rm Higgs}}_{ij} &=&
  \delta_{ig}\delta_{jg}\frac{\delta(1-x)}{1-\epsilon}
  \,,
\end{eqnarray}
and
\begin{eqnarray}
  \tilde\sigma^{(0),{\rm DY}}_{ij} &=&
  \delta_{iq}\delta_{j\bar{q}}\delta(1-x)(1-\epsilon)
  \,.
\end{eqnarray}
The indices $i$ and $j$ in Eq.~(\ref{eq::sigtil}) refer to the
partons in the initial state, i.e., gluons and massless quarks.

The crucial input to the convolution integrals are the splitting functions
$P_{ij}(x)$ describing the probability of parton $j$ to turn by
emission into parton $i$ with the fraction $x$ of its initial
energy. We define the perturbative expansion by
\begin{eqnarray}
  P_{ij}(x) &=& \delta_{ij} \delta(1 - x)
  + \frac{\alpha_s}{\pi} P_{ij}^{(1)}(x)
  + \left(\frac{\alpha_s}{\pi}\right)^2 P_{ij}^{(2)}(x)
  + \left(\frac{\alpha_s}{\pi}\right)^3 P_{ij}^{(3)}(x)
  + \ldots\,,
  \label{eq::Pij}
\end{eqnarray}
where the analytic results for the $k$-loop splitting function
$P_{ij}^{(k)}(x)$ can be found in
Refs.~\cite{Altarelli:1977zs,Curci:1980uw,Moch:2004pa,Vogt:2004mw}.
Together with the program \prog{MT} we provide a \prog{Mathematica} file
\prog{splitfnsMVV.m} which contains splitting functions computed
  in~\cite{Moch:2004pa,Vogt:2004mw} in a slightly modified form.
  Namely, the upper index is shifted by one and the singular
contributions are stated explicitly in terms of generalized functions.
In Tab.~\ref{tab::splitfnsMVV} we provide the translation from the
notation used in the file to that of Eq.~(\ref{eq::Pij}). Note that
for the quark contributions the indices $i$ and $j$ are omitted and
additional super- and subscripts are introduced in order to distinguish
the various contributions.

\begin{table}[t]
  \begin{center}
    \begin{tabular}{c|c|p{230 pt}}
      \hline
      \prog{splitfnsMVV.m}
      & Eq.~(\ref{eq::Pij})
      & comments, connection to \cite{Moch:2004pa,Vogt:2004mw} \\
      \hline
      \verb|Pgg1| & $P_{gg}^{(1)}$ & --- \\
      \verb|Pgq1| & $P_{gq}^{(1)}$ & --- \\
      \verb|Pqg1| & $P_{qg}^{(1)}$ & --- \\
      \verb|Pqq1| & $P_{\rm ns}^{(1)}$ & --- \\
      \verb|Pgg2| & $P_{gg}^{(2)}$ & --- \\
      \verb|Pgq2| & $P_{gq}^{(2)}$ & --- \\
      \verb|Pqg2| & $P_{qg}^{(2)}$ & --- \\
      \verb|Pnsp2| & $P_{\rm ns}^{(2)+}$ & quark-quark non-singlet, equals $P_{qq}^{\rm v}+P_{q\bar{q}}^{\rm v}$\\
      \verb|Pnsm2| & $P_{\rm ns}^{(2)-}$ & quark-quark non-singlet, equals $P_{qq}^{\rm v}-P_{q\bar{q}}^{\rm v}$\\
      \verb|Pps2| & $P_{\rm ps}^{(2)}$ & quark-quark pure singlet \\
      \verb|Pgg3| & $P_{gg}^{(3)}$ & --- \\
      \verb|Pgq3| & $P_{gq}^{(3)}$ & --- \\
     \verb|Pqg3| & $P_{qg}^{(3)}$ & --- \\
     \verb|Pnsp3| & $P_{\rm ns}^{(3)+}$ & quark-quark non-singlet, equals $P_{qq}^{\rm v}+P_{q\bar{q}}^{\rm v}$\\
     \verb|Pnsm3| & $P_{\rm ns}^{(3)-}+P_{\rm ns}^{(3) \rm s}$ &
     quark-quark non-singlet, equals $P_{qq}^{\rm v}-P_{q\bar{q}}^{\rm
       v}+P_{\rm ns}^{\rm s}$,\newline $P_{\rm ns}^{(3) \rm s}$
     is proportional to $d^{abc}d_{abc}/n_c$\\
     \verb|Pps3| & $P_{\rm ps}^{(3)}$ & quark-quark pure singlet \\
     \hline
    \end{tabular}
    \caption{\label{tab::splitfnsMVV} Notation for $P_{ij}^{(k)}$ used
      in \prog{splitfnsMVV.m} and in Eq.~(\ref{eq::Pij}).  The
      superscripts v and s stand for ``valence'' and ``sea''.  Note that
      the superscript indicating the loop-order is shifted in our
      notation compared to Refs.~\cite{Moch:2004pa,Vogt:2004mw} by $+1$
      and that $P_{\rm ns}^{(3) \rm s}$ appears at three loops for the
      first time.}
  \end{center}
\end{table}

The outline of the paper is as follows: the algorithm realized in
\prog{MT} is described in the following section, its functionality in
Section~\ref{sec::MT}.  Examples for the application of \prog{MT} are
given in Sections~\ref{sec::ggh} and~\ref{sec::DY} where the subtraction
terms are discussed for Higgs boson production and the Drell-Yan process
at N$^3$LO.  A summary is given in Section~\ref{sec::sum}.


\section{Systematic approach to convolution integrals}

Convolution integrals of partonic cross sections and splitting functions
are defined as
\begin{eqnarray}
  \left[f \otimes g \right](x) &=&
  \int_0^1 {\rm d}x_1 {\rm d}x_2 \delta(x - x_1 x_2) f(x_1) g(x_2)\,,
\end{eqnarray}
where the functions $f$ and $g$ include combinations of HPLs (for
the definition see below) up to
certain maximum weight with factors $1/x$, $1/(1-x)$, and $1/(1+x)$,
polynomials in $x$, and the generalized functions $\delta(1-x)$ and
$\left[\frac{\ln^k(1-x)}{1-x}\right]_+$.

Although the algorithm used for the computation of the convolution
integrals has already been presented before (see Appendix~B
of~\cite{Pak:2011hs} and Ref.~\cite{Hoschele:2012xc}), we decided to
review its description for completeness' sake.  An alternative
way to compute the same convolution integrals has been
presented in Ref.~\cite{Buehler:2013fha}.

Convolutions $[f \otimes g]$ are most conveniently dealt with by
considering Mellin transforms defined through
\begin{eqnarray}
  M_n\left[f(x)\right] &=& \int_0^1 {\rm d}x x^{n-1} f(x)\,,
  \label{eqn:mellinn}
\end{eqnarray}
since in Mellin space convolution integrals turn into products of Mellin
images:
\begin{eqnarray}
  M_n\big[ [f \otimes g](x) \big] &=&
  M_n\left[f(x)\right] M_n\left[g(x)\right]\,.
\end{eqnarray}

Mellin transforms relate HPLs and their
derivatives to harmonic sums~\cite{Vermaseren:1998uu,Remiddi:1999ew}.
HPLs $\HPL{\vec{a}}{x}$ are nested integrals,
recursively defined via
\begin{eqnarray}
  \HPL{}{x} = 1,~~
  \HPL{a,\vec{b}}{x} =
  \int_0^x {\rm d}x^\prime f_a(x^\prime)~ \HPL{\vec{b}}{x^\prime},~~
  f_0(x) = \frac{1}{x},~
  f_1(x) = \frac{1}{1 - x},~
  f_{-1}(x) = \frac{1}{1 + x}
  \,,
\end{eqnarray}
where one refers to the number of indices as weight.  Up to weight three
HPLs may be represented by combinations of ordinary logarithms and
Nielsen polylogarithms. E.g. up to weight one we have
\begin{eqnarray}
  \HPL{}{x} &=& 1\,, \nonumber\\
  \HPL{0}{x} &=& \ln{x}\,, \nonumber\\
  \HPL{1}{x} &=& -\ln(1-x)\,, \nonumber\\
  \HPL{-1}{x} &=& \ln(1+x)\,.
\end{eqnarray}
Harmonic sums $\HS{\vec{a}}{x}$ are defined in a similar fashion as nested sums:
\begin{eqnarray}
  \HS{}{n} = 1,~~
  \HS{a,\vec{b}}{n} = \sum_{i=1}^n f_a(i)~ \HS{\vec{b}}{i},~~
  f_a(i) = \left\{
    \begin{array}{ll}
      i^{-a}, &  a \ge 0\,, \\
      (-1)^i ~i^a, & a < 0\,,
    \end{array}\right.
  \label{eqn:hsums}
\end{eqnarray}
where the weight is defined as the sum of the absolute values of the indices.

For illustration we demonstrate Mellin transforms of HPLs through
weight one:
\begin{eqnarray}
  M_n[1] &=& \frac{1}{n}\,, \nonumber\\
  M_n[\HPL{0}{x}] &=& -\frac{1}{n^2}\,, \nonumber\\
  M_n[\HPL{1}{x}] &=& \frac{\HS{1}{n}}{n}\,, \nonumber\\
  M_n[\HPL{-1}{x}] &=& -\frac{(-1)^n}{n} \left( \HS{-1}{n} + \ln{2}
  \right) + \frac{\ln{2}}{n}\,.
  \label{eqn:mttab}
\end{eqnarray}

The main ingredient in our algorithm are relations among Mellin transforms
of functions $f(x)$, $x^k f(x)$ and $d f(x)/d x$ established with the help of
integration-by-parts identities. From the definition of the Mellin transform
it is obvious that
\begin{eqnarray}
  M_n\left[ x^k f(x)\right] &=& M_{n+k}\left[ f(x) \right]\,,
  \label{eqn:shift}
\end{eqnarray}
and furthermore if
$f(x)$ is regular for $x\to1$, then
\begin{eqnarray}
  M_n\left[ \frac{d}{dx} f(x) \right] &=&
  x^{n-1} f(x) \Big|_0^1 - (n-1) M_{n-1}\left[ f(x) \right]\,.
  \label{eqn:deriv}
\end{eqnarray}
Note that in the limit $x\to 0$ the boundary term in Eq.~\eqref{eqn:deriv}
vanishes since we always consider $n$ higher than the order of the
highest pole that $f(x)$ may have at $x=0$.
However, in the limit $x\to 1$ logarithmic singularities of the form
$\ln^k(1-x)$ may appear.
They are treated with the help of
``regularized derivatives'' $\hat{\partial}_x$ which are defined as
\begin{eqnarray}
  M_n\left[ \hat{\partial}_x 1 \right] &=& 1\,, \nonumber\\
  M_n\left[ \hat{\partial}_x f(x) \right] &=&
  R\left[f(x)\right] - (n-1) M_{n-1}\left[ f(x) \right]\,.
  \label{eqn:rderiv}
\end{eqnarray}
The operation $R$ regulates the boundary term by dropping all
logarithmically divergent contributions:
\begin{eqnarray}
  R[g_k(x)\ln^k(1-x) + g_{k-1}(x) \ln^{k-1}(1-x) + \ldots + g_0(x)] = g_0(1)\,,~~
  g_k(1) \ne 0~~ \forall k > 0\,.
\end{eqnarray}

To motivate this definition let us remark that
it corresponds to regulating divergences by means of
delta and plus distributions. As an example consider
$f(x)=\HPL{1}{x}=-\ln(1-x)$. Since $d \HPL{1}{x} / d x = 1/(1-x)$
the Mellin transform
$M_n[ d f(x) / d x ]$
is not defined.
On the other hand, if we replace $1/(1-x)$ by the corresponding
plus distribution we have
\begin{eqnarray}
  M_n\left[ \left[\frac{1}{1-x}\right]_+ \right] &=& -\HS{1}{n-1}
  \,.
\end{eqnarray}
The same result is obtained by applying Eq.~\eqref{eqn:rderiv} as can be seen by
the following chain of equations
\begin{eqnarray}
  M_n\left[ \hat{\partial}_x \HPL{1}{x} \right]
  = R\left[ -\ln(1-x) \right] - (n-1) M_{n-1}\left[ -\ln(1-x) \right]
  = 0 - (n-1) \frac{\HS{1}{n-1}}{n-1} = - \HS{1}{n-1}
  \,.
  \nonumber
\end{eqnarray}

Furthermore, in case $f(x)$ is not divergent for $x\to1$
Eq.~\eqref{eqn:rderiv} reduces to the usual derivative
(see Eq.~\eqref{eqn:deriv}).
For example,
\begin{eqnarray}
  \hat{\partial}_x \HPL{-1}{x} =
  \frac{d}{dx} \HPL{-1}{x} = \frac{1}{1+x}\,,~~
  \hat{\partial}_x \HPL{-1,1}{x} =
  \frac{d}{dx} \HPL{-1,1}{x} = \frac{\HPL{1}{x}}{1+x}\,.
\end{eqnarray}

The concept of regularized derivatives relates HPLs to ``common''
generalized functions.  This allows for a unified treatment of Mellin
transforms of derivatives of HPLs, independent of the presence of
divergences. Hence it is sufficient to consider Mellin transforms of
regularized derivatives in case Mellin transforms for generalized
functions are needed.  To illustrate this point let us show a few
examples for regularized derivatives of HPLs
\begin{eqnarray}
  \hat{\partial}_x~ 1 &=& \delta(1 - x)\,, \nonumber\\
  \hat{\partial}_x \HPL{1}{x} &=&
  \left[\frac{1}{1-x}\right]_+, \nonumber\\
  \hat{\partial}_x \HPL{1,1}{x} &=&
  -\left[\frac{\ln(1-x)}{1-x}\right]_+, \nonumber\\
  \hat{\partial}_x \HPL{1,1,1}{x} &=&
  \frac{1}{2} \left[\frac{\ln^2(1-x)}{1-x}\right]_+, \nonumber\\
  \hat{\partial}_x \HPL{1,2}{x} &=&
  \frac{\pi^2}{6} \left[\frac{1}{1-x}\right]_+ + \frac{\HPL{2}{x} - \frac{\pi^2}{6}}{1-x}\,.
\end{eqnarray}

To continue the example of HPLs up to weight one, we list the Mellin
transforms of their regularized derivatives
\begin{eqnarray}
  M_n[\hat{\partial}_x 1] &=& 1\,, \nonumber\\
  M_n[\hat{\partial}_x \HPL{0}{x}] &=& \frac{1}{n-1}\,, \nonumber\\
  M_n[\hat{\partial}_x \HPL{1}{x}] &=& -\HS{1}{n-1}\,, \nonumber\\
  M_n[\hat{\partial}_x \HPL{-1}{x}] &=&
  (-1)^{n-1}\HS{-1}{n-1} + (-1)^{n-1}\ln{2}\,.
  \label{eq::MTgenderiv}
\end{eqnarray}

One can interpret the combination of Mellin transforms of HPLs (see
Eq.~\eqref{eqn:mttab}) and their generalized derivatives (see
Eq.~\eqref{eq::MTgenderiv}) computed up to a fixed maximum weight as a
system of linear equations.  This system can be solved for monomials of
the form $1/n^k$, $\HS{\vec{a}}{n}/n^k$, and $(-1)^n
\HS{\vec{a}}{n}/n^k$.  The solution is then equivalent to the
  inverse Mellin transform.

We conclude our example for quantities up to weight one by providing
their inverse Mellin transforms:
\begin{eqnarray}
  \MI{x}{\frac{1}{n}} &=& 1\,,\nonumber\\
  \MI{x}{1} &=& \regd{x} 1\,,\nonumber\\
  \MI{x}{\frac{1}{n^2}} &=& -\HPL{0}{x}\,,\nonumber\\
  \MI{x}{S_1(n)} &=& -x \regd{x} \HPL{1}{x}\,,\nonumber\\
  \MI{x}{\frac{S_1(n)}{n}} &=& \HPL{1}{x}\,.
\end{eqnarray}

The discussion above allows us to formulate the following algorithm.
Given the task of finding a convolution of functions f and g, one may perform
the following steps.
\begin{enumerate}
\item Transform the expressions $f$ and $g$ to Mellin $n$-space.
\item Compute a table of Mellin transforms of HPLs up to a fixed maximum
  weight.\label{enum::tab1}
\item Prepare a corresponding table holding regularized derivatives
  of HPLs.\label{enum::tab2}
\item Solve the system of linear equations composed of the tables from
  steps~\ref{enum::tab1} and~\ref{enum::tab2}. \label{enum::sol}
\item Perform inverse transformation to $x$-space by substituting
    the results from step~\ref{enum::sol} into the expression
    $\MI{n}{\MT{n}{f}\MT{n}{g}}$.
\end{enumerate}
In step 2, Mellin transforms of HPLs are obtained by using
the \prog{FORM} package \prog{harmpol}~\cite{harmpol}.
In step 3, Mellin transforms of regularized derivatives of HPLs
are obtained from the result of step 2 via Eq.~\eqref{eqn:rderiv}.
Although some inverse transforms cannot be determined from the system
of equations in step 4, in practice they cancel in final expressions
for the convolutions in step 5.
In \prog{MT} the precomputed results of steps 2 and 4 are tabulated
to speed-up the calculation and provided in the form of table files.

In the considered examples of Higgs boson and vector boson production
(cf. Sections~\ref{sec::ggh} and~\ref{sec::DY}) we were able to compute all
necessary convolutions relevant at N$^3$LO applying this algorithm.


\section{\label{sec::MT}
  Description of \prog{MT}}

\prog{MT} is a \prog{Mathematica} package for computing
analytically convolutions of HPLs with factors $1/x$, $1/(1-x)$ and
$1/(1+x)$, polynomials in $x$, and generalized functions $\delta(1-x)$
and $\left[ \frac{\ln^k(1-x)}{1-x} \right]_+$, via Mellin
transforms.  For cross checks it also provides capability to compute
  convolutions numerically.  It requires \prog{Mathematica}
version 6 or later and the \prog{HPL}
package~\cite{Maitre:2005uu,Maitre:2007kp}.  Having installed \prog{MT}
properly as described in the provided \prog{README} file, one can load
it via
\begin{mmacode}
\In
\begin{minipage}[t]{\codewidth}
\begin{verbatim}
<<MT`
\end{verbatim}
\end{minipage}
\end{mmacode}
The package handles the following objects:
\begin{itemize}
\item
\cmd{HPL[\{$m_1$,$\ldots$,$m_N$\}, $x$]} $= \H_{m_1,\ldots,m_N}(x)$ is the HPL
provided by \prog{HPL} package.

\item
\cmd{PlusDistribution[$k$, $1-x$]} = $\left[ \frac{\ln^k(1-x)}{1-x}\right]_+$
is the plus distribution ($k = 0, 1, 2, \ldots$). \\
\cmd{PlusDistrubition[$-1$, $1-x$]} = $\delta(1-x)$ is the delta function.

\item
\cmd{DReg[$f$, $x$]} $ = \hat\partial_x f$ is the regularized derivative of
the expression $f$ with respect to $x$ which can be obtained by replacing
every singular term $\frac{\ln^k(1-x)}{1-x}$ in the result of the ordinary
differentiation $\frac{df}{dx}$ with a plus distribution
$\left[ \frac{\ln^k(1-x)}{1-x}\right]_+$.
We define \cmd{DReg[1, $x$]} $ = \hat\partial_x 1 = \delta(1-x)$.
Mellin transforms of regularized derivatives are given by Eq.~\eqref{eqn:rderiv}.

\item
\cmd{HSum[\{$m_1$,$\ldots$,$m_N$\}, $n$]} $= \S_{m_1,\ldots,m_N}(n)$ is the
harmonic sum which appears in Mellin transforms of
HPLs.
\end{itemize}

For computing convolutions, both analytically and numerically, \prog{MT} package
provides the following functions:
\begin{itemize}
\item
\cmd{Convolution[$f_1$,$\ldots$,$f_N$,$x$]} gives the convolution with
respect to $x$ of the expressions $f_1$, $\ldots$, $f_N$,
\begin{equation}
  [f_1 \otimes \ldots \otimes f_N](x) =
  \int_0^1 dx_1 \ldots \int_0^1 dx_N \, f_1(x_1) \ldots f_N(x_N)
  \delta(x - x_1 \ldots x_N).
\end{equation}
For illustration we consider the following examples:
\begin{mmacode}
\In
\begin{minipage}[t]{\codewidth}%
\begin{verbatim}
Convolution[HPL[{0},x]/(1+x),HPL[{0},x]/(1-x),x]
\end{verbatim}
\end{minipage}
\\
\Out
$ \displaystyle
\frac{\H_{-2,0}(x)}{x+1}
- \frac{\H_{2,0}(x)}{x+1}
- \frac{\H_{0,0,0}(x)}{x+1}
- \frac{\pi^2\H_{0}(x)}{12(x+1)}
- \frac{\zeta(3)}{2(x+1)}
$
\\
\In
\begin{minipage}[t]{\codewidth}%
\begin{verbatim}
Convolution[PlusDistribution[0,1-x],(HPL[{2},x]-HPL[{2},1])/(1-x),x]
\end{verbatim}
\end{minipage}
\\
\Out
$ \displaystyle
- \frac{\H_{1,2}(x)}{1-x}
- \frac{\H_{2,0}(x)}{1-x}
- \frac{2\H_{2,1}(x)}{1-x}
+ \frac{\pi^2\H_{0}(x)}{6(1-x)}
+ \frac{\pi^2\H_{1}(x)}{6(1-x)}
- \frac{\H_{3}(x)}{1-x}
- \frac{\zeta(3)}{1-x}
$
\\
\In
\begin{minipage}[t]{\codewidth}%
\begin{verbatim}
Convolution[PlusDistribution[0,1-x],PlusDistribution[0,1-x],x]
\end{verbatim}
\end{minipage}
\\
\Out
$ \displaystyle
2 \left[ \frac{\log(1-x)}{1-x} \right]_+
- \frac{\H_0(x)}{1-x}
- \frac{1}{6} \pi^2 \delta(1-x)
$
\end{mmacode}
Since \cmd{Convolution} converts the convolution into a product of
Mellin transforms of the expressions and then performs the inverse
Mellin transform to obtain the result, the maximum weight of the expressions
that \cmd{Convolution} can handle is limited by the weight implemented
in the tables of the transforms. In the current version
\prog{MT} can handle expressions up to  weight six by
default,\footnote{In the final result for the N$^3$LO contributions only
  weight-five HPLs appear. However, in intermediate steps harmonic sums
  of weight six are present which require the corresponding
  tables.} tables up to weight eight can be loaded if necessary.
In case higher weights are needed \cmd{Convolution} returns
a result containing unevaluated Mellin and inverse Mellin transforms.

\item \cmd{NConvolution[$f_1$,$\ldots$,$f_N$,$x$,$a$]} numerically computes
  the convolution with respect to $x$ of the expressions $f_1$, $\ldots$,
  $f_N$ at $x = a$, by using \prog{Mathematica}'s built-in \cmd{NIntegrate} function.
  In case the result contains plus distributions \cmd{NConvolution} evaluates
  the corresponding coefficients numerically.  The numerical computation of
  the above examples for $x=3/10$ looks as follows:
\begin{mmacode}
\In
\begin{minipage}[t]{\codewidth}%
\begin{verbatim}
NConvolution[HPL[{0},x]/(1+x),HPL[{0},x]/(1-x),x,3/10]
\end{verbatim}
\end{minipage}
\\
\Out
$ \displaystyle
0.600799
$
\\
\In
\begin{minipage}[t]{\codewidth}%
\begin{verbatim}
NConvolution[PlusDistribution[0,1-x],(HPL[{2},x]-HPL[{2},1])/(1-x),x,3/10]
\end{verbatim}
\end{minipage}
\\
\Out
$ \displaystyle
-2.86725+1.7975\times10^{-17} \ii
$
\\
\In
\begin{minipage}[t]{\codewidth}%
\begin{verbatim}
NConvolution[PlusDistribution[0,1-x],PlusDistribution[0,1-x],x,3/10]
\end{verbatim}
\end{minipage}
\\
\Out
$ \displaystyle
2. \left[ \frac{\log(1-x)}{1-x} \right]_+
- 1.64493 \, \delta(1-x)
+ 1.71996
$
\end{mmacode}

Currently \cmd{NConvolution} supports convolutions of up to three functions
exclusive of delta functions, which can be removed trivially.
If the expressions contain plus distributions the auxiliary
regularization~\cite{Anastasiou:2005qj}
\begin{equation}
  \frac{\ln^k(1-x)}{1-x} \to \lim_{a\to1}
  \frac{1}{\eta^k} \frac{\partial^k}{\partial a^k} (1-x)^{-1+a \eta}
\end{equation}
is introduced before the numerical integration of each individual
term. In a next step the singularities in the integrals for $x\to0$ 
and $x\to1$ are separated, 
the differentiation w.r.t. $a$ is performed,
the limit $a\to1$ is taken, and the result is expanded in $\eta$.
All pole terms $1/\eta^k$ should cancel among the integrals and the
leading term gives the result of the convolution. The remaining
integrals must have no divergences and can be performed
numerically.  \cmd{NConvolution} checks that all pole terms cancel
  within an absolute tolerance controlled by \cmd{Tolerance} option
($10^{-6}$ by default).

Since \cmd{NConvolution} performs the convolution numerically, the
result must be finite. Moreover it must be possible to evaluate the
integrand numerically, i.e., it cannot contain any symbols, except those
specified by \cmd{Constants} option. If \cmd{Constants} option is used,
\cmd{NConvolution} tries to collect terms involving the same powers of
these constants and splits the convolution with respect to them.  The
use of \cmd{Constants} is illustrated by the following example:
\begin{mmacode}
\In
\begin{minipage}[t]{\codewidth}%
\begin{verbatim}
NConvolution[(1+c)*x+x^2,HPL[{0},x],x,3/10,Constants->{c}]
\end{verbatim}
\end{minipage}
\\
\Out
$ \displaystyle
-0.503973 \, c -0.878459
$
\end{mmacode}
\cmd{NConvolution} accepts also the following options, which are passed to
\cmd{NIntegrate}: \cmd{AccuracyGoal}, \cmd{MaxPoints}, \cmd{MaxRecursion},
\cmd{Method}, \cmd{MinRecursion}, \cmd{PrecisionGoal}, and
\cmd{WorkingPrecision}.
One may need to change these options when the convergence of the numerical
integration is slow. Note that \cmd{NConvolution} may turn out to be very slow
if it is called with more than two functions in the argument.

\item
\cmd{NEval[$f$, $x$, $a$]} gives the numerical value of the expression $f$
at $x=a$ and is similar to \cmd{N[$f$ /. \!\!\!$x$ $\to$ $a$]} but does not
touch $x$ in generalized functions.
This is convenient in case one wants to obtain a numerical value by
substituting a certain
value of $x$ into a result of \cmd{Convolution} which contains plus
distributions. For example the comparison with \cmd{NConvolution}
(see example above) may look as follows:
\begin{mmacode}
\In
\begin{minipage}[t]{\codewidth}%
\begin{verbatim}
Convolution[PlusDistribution[0,1-x],PlusDistribution[0,1-x],x]
\end{verbatim}
\end{minipage}
\\
\Out
$ \displaystyle
2 \left[ \frac{\log(1-x)}{1-x} \right]_+
- \frac{\H_0(x)}{1-x}
- \frac{1}{6} \pi^2 \delta(1-x)
$
\\
\In
\begin{minipage}[t]{\codewidth}%
\begin{verbatim}
NEval[%,x,3/10]
\end{verbatim}
\end{minipage}
\\
\Out
$ \displaystyle
2. \left[ \frac{\log(1-x)}{1-x} \right]_+
- 1.64493 \, \delta(1-x)
+ 1.71996
$
\end{mmacode}
\cmd{NEval[$f$, $x$, $a$, $n$]} attempts to output the result with
$n$-digit precision.

\end{itemize}

\cmd{Convolution} utilizes Mellin transforms and inverse Mellin
transforms for computing convolutions. \prog{MT} package also provides
functions that allow users to find Mellin transforms and inverse Mellin
transforms:
\begin{itemize}

\item
\cmd{MTMellinn[$f$, $x$, $n$]} computes the Mellin transform with respect to $x$
of the expression $f$:
\begin{equation}
  M_n \bigl[ f(x) \bigr] = \int_0^1 dx \, x^{n-1} f(x)\,,
\end{equation}
which is illustrated in the following example:
\begin{mmacode}
\In
\begin{minipage}[t]{\codewidth}%
\begin{verbatim}
MTMellinn[HPL[{1},x]/(1+x),x,n]
\end{verbatim}
\end{minipage}
\\
\Out
$ \displaystyle
(-1)^{n-1} \S_{-1,1}(n-1)
+ \frac{1}{12} \pi^2 (-1)^{n-1}
- \frac{1}{2} (-1)^{n-1} \log^2(2)
$
\\
\In
\begin{minipage}[t]{\codewidth}%
\begin{verbatim}
MTMellinn[(HPL[{2},x]-HPL[{2},1])/(1-x),x,n]
\end{verbatim}
\end{minipage}
\\
\Out
$ \displaystyle
\S_{2,1}(n-1)
- 2 \zeta(3)
$
\\
\In
\begin{minipage}[t]{\codewidth}%
\begin{verbatim}
MTMellinn[PlusDistribution[2,1-x],x,n]
\end{verbatim}
\end{minipage}
\\
\Out
$ \displaystyle
-2\S_{1,1,1}(n-1)
$
\end{mmacode}

\item
\cmd{NMTMellinn[$f$, $x$, $a$]} numerically computes the Mellin transform
with respect to $x$ of the expression $f$ with $n=a$.
\begin{mmacode}
\In
\begin{minipage}[t]{\codewidth}%
\begin{verbatim}
NMTMellinn[HPL[{1},x]/(1+x),x,12]
\end{verbatim}
\end{minipage}
\\
\Out
$ \displaystyle
0.133014
$
\\
\In
\begin{minipage}[t]{\codewidth}%
\begin{verbatim}
NMTMellinn[(HPL[{2},x]-HPL[{2},1])/(1-x),x,12]
\end{verbatim}
\end{minipage}
\\
\Out
$ \displaystyle
-0.351258
$
\\
\In
\begin{minipage}[t]{\codewidth}%
\begin{verbatim}
NMTMellinn[PlusDistribution[2,1-x],x,12]
\end{verbatim}
\end{minipage}
\\
\Out
$ \displaystyle
-14.684
$
\end{mmacode}
\cmd{NMTMellinn} accepts the following options (see also \cmd{NConvolution}):
\cmd{AccuracyGoal}, \cmd{Constants}, \cmd{MaxPoints}, \cmd{MaxRecursion},
\cmd{Method}, \cmd{MinRecursion}, \cmd{PrecisionGoal}, and
\cmd{WorkingPrecision}.

\item
\cmd{MTInverse[$f$, $x$, $n$]} gives the inverse Mellin transform with
respect to $n$ of the expression $f$.
\begin{mmacode}
\In
\begin{minipage}[t]{\codewidth}%
\begin{verbatim}
MTInverse[HSum[{2,1},n]/n,x,n]
\end{verbatim}
\end{minipage}
\\
\Out
$ \displaystyle
\frac{1}{6} \pi^2 \H_1(x) - \H_{1,2}(x)
$
\\
\In
\begin{minipage}[t]{\codewidth}%
\begin{verbatim}
MTInverse[(-1)^n*HSum[{-1},n]/n,x,n]
\end{verbatim}
\end{minipage}
\\
\Out
$ \displaystyle
-\log(2) \, \cmd{MTInverse}\left(\frac{(-1)^n}{n},x,n\right)
- \H_{-1}(x)
+ \log(2)
$
\end{mmacode}
In the latter example the inverse Mellin transform cannot be
  expressed in terms of HPLs and is instead reduced to a simpler one.

\end{itemize}

In the following we list further functions defined in \prog{MT} package
which may be useful for users:
\begin{itemize}
\item
\cmd{MTPlusToDReg[$f$]} converts plus distributions in the expression $f$
into regularized derivatives.
\begin{mmacode}
\In
\begin{minipage}[t]{\codewidth}%
\begin{verbatim}
MTPlusToDReg[PlusDistribution[3,1-x]]
\end{verbatim}
\end{minipage}
\\
\Out
$ \displaystyle
-6 \, \hat\partial_x \H_{1,1,1,1}(x)
$
\end{mmacode}

\item
\cmd{MTDRegToPlus[$f$]} converts regularized derivatives in the expression
$f$ into plus distributions.
\begin{mmacode}
\In
\begin{minipage}[t]{\codewidth}%
\begin{verbatim}
MTDRegToPlus[DReg[HPL[{1,1,1,1},x],x]]
\end{verbatim}
\end{minipage}
\\
\Out
$ \displaystyle
-\frac{1}{6} \left[ \frac{\log^3(1-x)}{1-x} \right]_+
$
\end{mmacode}

\item
\cmd{MTPlusSimplify[$f$]} performs transformations on terms in the
expression $f$ containing generalized functions such that their
coefficients become constants:
\begin{align}
  \delta(1-x) f(x) &= \delta(1-x) f(1) , \\
  \left[ \frac{\ln^k(1-x)}{1-x} \right]_+ f(x)
  &= \left[ \frac{\ln^k(1-x)}{1-x} \right]_+ f(1)
  + \frac{\ln^k(1-x)}{1-x} \bigl[ f(x) - f(1) \bigr] .
\end{align}
\begin{mmacode}
\In
\begin{minipage}[t]{\codewidth}%
\begin{verbatim}
MTPlusSimplify[PlusDistribution[1,1-x]*HPL[{2},x]]
\end{verbatim}
\end{minipage}
\\
\Out
$ \displaystyle
\frac{1}{6} \pi^2 \left[ \frac{\log(1-x)}{1-x} \right]_+
+ \frac{\H_1(x) \bigl( \pi^2 - 6 \H_2(x) \bigr)}{6(1-x)}
$
\end{mmacode}

\item
\cmd{MTHarmonize[$f$, $n$]} shifts the argument of harmonic sums in the
expression $f$ such that each term in the result individually
depends on a single argument $n + i$.
\begin{mmacode}
\In
\begin{minipage}[t]{\codewidth}%
\begin{verbatim}
MTHarmonize[HSum[{1,-1},n+2]/n,n]
\end{verbatim}
\end{minipage}
\\
\Out
$ \displaystyle
\frac{\S_{1,-1}(n)}{n}
+\frac{3\S_{-1}(n)}{2n}
-\frac{\S_{-1}(n+1)}{n+1}
-\frac{\S_{-1}(n+2)}{2(n+2)}
-\frac{5(-1)^n}{4n}
-\frac{3(-1)^{n+1}}{2(n+1)}
-\frac{(-1)^{n+2}}{4(n+2)}
$
\end{mmacode}

\item
\cmd{MTNormalize[$f$, $n$]} normalizes the argument of harmonic sums in the
expression $f$ such that they have $n$ as argument.
\begin{mmacode}
\In
\begin{minipage}[t]{\codewidth}%
\begin{verbatim}
MTNormalize[HSum[{-1,2},n+1],n]
\end{verbatim}
\end{minipage}
\\
\Out
$ \displaystyle
\S_{-1,2}(n)
- \frac{(-1)^n \S_2(n)}{n+1}
- \frac{(-1)^n}{(n+1)^3}
$
\end{mmacode}

\item
\cmd{MTProductExpand[$f$]} substitutes products of harmonic sums in the
expression $f$ with harmonic sums of higher weights. For a
  discussion of the corresponding algebra see, e.g., Appendix~B
of~\cite{Pak:2011hs}.
\begin{mmacode}
\In
\begin{minipage}[t]{\codewidth}%
\begin{verbatim}
MTProductExpand[HSum[{2},n]*HSum[{-1},n]]
\end{verbatim}
\end{minipage}
\\
\Out
$ \displaystyle
\S_{-1,2}(n)
+\S_{2,-1}(n)
-\S_{-3}(n)
$
\end{mmacode}

\item
\cmd{CancelMTInverse[$f$]} tries to cancel unevaluated inverse Mellin transforms
in the expression $f$.
\begin{mmacode}
\In
\begin{minipage}[t]{\codewidth}%
\begin{verbatim}
MTInverse[HSum[{-1},n+1],x,n] - MTInverse[HSum[{-1},n] - (-1)^n/(n+1),x,n]
\end{verbatim}
\end{minipage}
\\
\Out
$ \displaystyle
x \, \cmd{MTInverse}\bigl(\S_{-1}(n),x,n\bigr)
- \cmd{MTInverse}\bigl(\S_{-1}(n),x,n\bigr)
- x \, \cmd{MTInverse}\left( \frac{(-1)^n}{n}, x, n \right)
$
\\
\In
\begin{minipage}[t]{\codewidth}%
\begin{verbatim}
CancelMTInverse[%]
\end{verbatim}
\end{minipage}
\\
\Out
$ \displaystyle
0
$
\end{mmacode}

\end{itemize}


\section{\label{sec::ggh} Example 1: collinear singularities for Higgs
  boson production at the LHC}

After the discovery of a new Higgs boson-like particle at the LHC it is now
of primary importance to measure its
properties like cross sections, branching ratios and couplings
with high precision. Current experimental results are
compatible with a Standard Model Higgs boson.

The main production mechanism for the latter is the gluon fusion
process. Although even second order corrections have been computed for
this process (see Refs.~\cite{Dittmaier:2011ti,Dittmaier:2012vm} and
references therein) the perturbative uncertainties are still of the
order of 10\% and thus it would be desirable to have the next term
in the perturbative expansion.  \prog{MT} can compute the
related subtraction terms for the collinear divergences originating
from radiation of partons off initial-state particles.

The N$^3$LO convolutions to this process have already been discussed in
Refs.~\cite{Hoschele:2012xc,Buehler:2013fha} where results for
  all convolutions are provided in electronic form. Thus let us at this
point only provide an example which demonstrates the use of \prog{MT}.

The results for the partonic cross sections up to NNLO expanded
sufficiently deeply in $\epsilon$ such that finite results at
N$^3$LO can be obtained are listed in the file
\prog{sig\_tilde\_LO\_NLO\_NNLO.m} which can be found on the
webpage~\cite{progdata_higgs} (see also Ref.~\cite{Hoschele:2012xc}).
As an example we consider the convolution of $\tilde{\sigma}^{(1)}_{qg}/x$
with $P_{gg}^{(1)}$ and $P_{\rm ns}^{(1)}$ which is computed as\footnote{In
  the examples we present in this and the next Section we do not show
  the complete output but abbreviate it using ellipses.}
\begin{mmacode}\setcounter{mmano}{0}
\In
\begin{minipage}[t]{\codewidth}%
\begin{verbatim}
<<MT`
\end{verbatim}
\end{minipage}
\\
\In
\begin{minipage}[t]{\codewidth}%
\begin{verbatim}
<<sig_tilde_LO_NLO_NNLO.m
\end{verbatim}
\end{minipage}
\\
\In
\begin{minipage}[t]{\codewidth}%
\begin{verbatim}
<<splitfnsMVV.m
\end{verbatim}
\end{minipage}
\\
\In
\begin{minipage}[t]{\codewidth}%
\begin{verbatim}
Convolution[rsigc["NLO", "qg"]/x, Pgg1/.splitfnsMVV, Pqq1/.splitfnsMVV, x];
\end{verbatim}
\end{minipage}
\\
\In
\begin{minipage}[t]{\codewidth}%
\begin{verbatim}
Collect[%, {ep, HPL[__], PlusDistribution[__], nl}, Together]
\end{verbatim}
\end{minipage}
\\
\Out
$ \displaystyle
\ldots + \bigg(\frac{4 (8 x^3+11 x^2+56 x-49)}{9 x}-\frac{8 {\rm nl} (x^2-2
x+2)}{27 x}\bigg) \H_{1,0}(x)+\ldots\newline
+{\rm ep}\bigg(\ldots +\H_{2}(x) \bigg(-\frac{4 (5 x^2+2 x+4)
  \H_{0,0}(\frac{{\rm
mu}^2}{{\rm Mh}^2})}{3 x}-\frac{2 {\rm nl} (9 x^2-26 x+7)}{27 x}\newline
+\frac{256 x^3+237 \pi^2 x^2+432 x^2-78 \pi^2 x+2832 x+276 \pi^2-1129}{27
x}\bigg)+\ldots\bigg)\newline
+{\rm ep}^2\bigg(\ldots-\frac{4 (9 x^2+6 x+4) \H_{5}(x)}{3
x}+\ldots+\frac{1}{19440 x} (-276480 x^3 \zeta(3)+2688 \pi^4
x^3+\ldots\newline-1538205)
\bigg)
$
\end{mmacode}
Note that the number of massless quarks both in
\prog{sig\_tilde\_LO\_NLO\_NNLO.m} and \prog{splitfnsMVV} is denoted by $n_l$.

In analogy to this example it is possible to obtain all convolution
contributions for the Higgs boson production by combining results from
\prog{sig\_tilde\_LO\_NLO\_NNLO.m} and \prog{splitfnsMVV.m}, including
multiple convolutions.


\section{\label{sec::DY}
  Example 2: collinear singularities for Drell-Yan production}

The Drell-Yan process, i.e., the production of a lepton pair in hadronic
collisions mediated by a vector
boson, constitutes an important benchmark process at hadron
colliders. In particular, it provides information about the PDFs and is
a useful tool in searches for heavier gauge bosons by examining the
invariant mass of the produced leptons.

In the following we briefly discuss the computation of the total cross
section up to NNLO with the emphasis on higher-order terms in
$\epsilon$. The results are later used in order to evaluate the
collinear subtraction terms at N$^3$LO using \prog{MT}.  In our
discussion we discard all contributions which contain a one-loop
triangle diagram with two gluons and a gauge boson as external
particles as subgraph since they cancel after summing over all
quarks of a family. Thus, we can restrict ourselves to the use of naive
anti-commuting $\gamma_5$ in the axial-vector coupling of the $W$ or $Z$
boson.

\begin{figure}[t]
  \begin{center}
    \begin{subfigure}[c]{0.25\textwidth}
      \centering
      \includegraphics{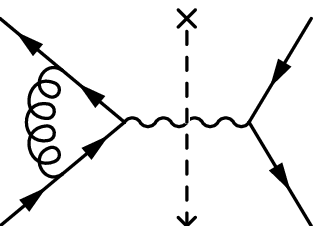}
      \vspace*{0.3pc}
      \caption{LO: $q\bar{q}$, v}
    \end{subfigure}
    \begin{subfigure}[c]{0.35\textwidth}
      \centering
      \includegraphics{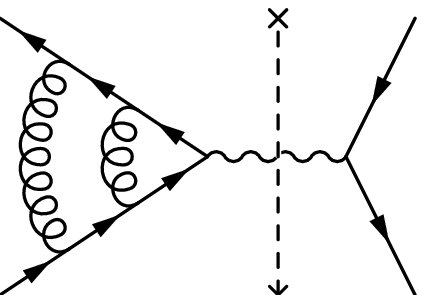}
      \vspace*{0.3pc}
      \caption{NLO: $q\bar{q}$, v}
    \end{subfigure}
    \begin{subfigure}[c]{0.35\textwidth}
      \centering
      \includegraphics{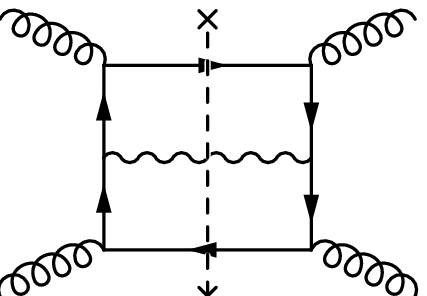}
      \vspace*{0.3pc}
      \caption{NNLO: $gg$, r$^2$}
    \end{subfigure}
    \vspace{1.0pc}\\
    \begin{subfigure}[c]{0.25\textwidth}
      \centering
      \includegraphics{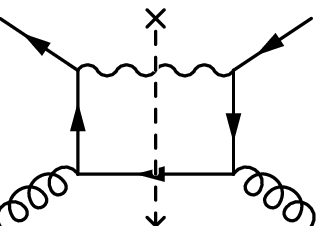}
      \vspace*{0.3pc}
      \caption{LO: $qg$, r}
    \end{subfigure}
    \begin{subfigure}[c]{0.3\textwidth}
      \centering
      \includegraphics{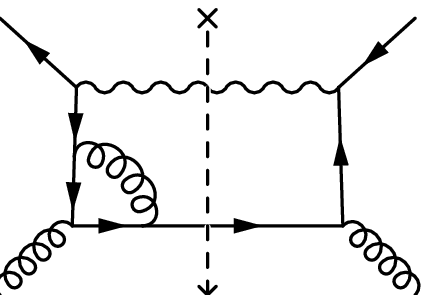}
      \vspace*{0.5pc}
      \caption{NLO: $qg$, r-v}
    \end{subfigure}
    \begin{subfigure}[c]{0.4\textwidth}
      \centering
      \includegraphics{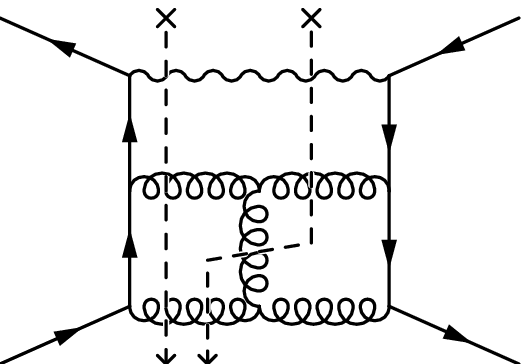}
      \vspace*{0.5pc}
      \caption{N$^3$LO: $q\bar{q}$, r$^2$-v and r$^3$}
    \end{subfigure}
    \caption{
      \label{fig::diag_DY}
      Sample Feynman diagrams up to N$^3$LO contributing to the
      Drell-Yan process.  The wiggly lines denote a generic vector
      boson, i.e., $\gamma$, $W$, or $Z$.  The dashed lines attached
      to crosses mark the considered cut(s) through a massive vector
      boson and additional massless particles.  The captions state the
      perturbative order, the channel and the type of contributions
      (``r'' for real, ``v'' for virtual, or their interference).
    }
  \end{center}
\end{figure}

For the calculation of the total cross section we apply the same techniques as
for the Higgs production~\cite{Anastasiou:2002yz,Pak:2009dg}, i.e., we use the
optical theorem and consider the imaginary part of the forward-scattering
amplitude by cutting the gauge boson, light quark and gluon lines.  It is
convenient to treat the purely virtual corrections separately since in that
case only the cut through the gauge boson is considered, see also
Fig.~\ref{fig::diag_DY}. The NNLO expression for the resulting quark form
factor can be found in
Refs.~\cite{Kramer:1986sg,Matsuura:1987wt,Matsuura:1988sm,Gehrmann:2005pd},
N$^3$LO results are presented in
Refs.~\cite{Baikov:2009bg,Gehrmann:2010ue,Lee:2010cga}.

Sample Feynman diagrams for the single- and double-real corrections,
considered simultaneously, can be found in
Fig.~\ref{fig::diag_DY}. At NLO and NNLO the following channels have to
be considered
\begin{itemize}
\item NLO: $q\bar{q}$ and $qg$,
\item NNLO: $q\bar{q}$, $qg$, $gg$, $qq$, $qq^\prime$,
\end{itemize}
where $q^\prime$ denotes a quark different from $q$ and the channels
involving $\bar{q}$ are not listed separately.

The first step in the evaluation of the Feynman integrals is the reduction to
master integrals using the Laporta algorithm~\cite{Laporta:2001dd}. We use our
own \prog{C++} 
implementation for this step. The NNLO master integrals are taken from
Ref.~\cite{Pak:2011hs} (see also~\cite{Anastasiou:2012kq}) which guarantees
that the final result is available including ${\cal O}(\epsilon)$ terms.

We present our result using the same notation as in
Ref.~\cite{Hamberg:1990np}, but set equal the renormalization and the
factorization scale, which does not reduce the complexity of the
convolutions to be calculated. In particular, we provide terms up to ${\cal
  O}(\epsilon^2)$ for the NLO quantities $\Delta_{q\bar q}^{(1)},
\Delta_{qg}^{(1)}$ of Eqs.~(B.2), (B.17) and up to ${\cal O}(\epsilon)$ for
the NNLO results in Eqs.~(B.7), (B.18), (B.21), (B.22),
(B.24)-(B.26).  For $\epsilon=0$ our results agree with
Refs.~\cite{Hamberg:1990np,Harlander:2002wh}.  We note that, in order to find
agreement we had to set the color factor $T_F=1/2$ in some cases.  Moreover in
the case of the gluon-gluon contribution the color factor $C_A C_F/T_F$ has
been attached to the labels $\verb|coeff[]|$ that will be discussed below.  We
refrain from explicitly listing analytic results in the paper but provide
computer-readable expressions in Ref.~\cite{progdata} (see also Appendix).

It is convenient to identify our partonic cross sections with results of
Eq.~(A.20) from Ref.~\cite{Hamberg:1990np} which provides a complete
expression for the hadronic cross section. The partonic results are easily
obtained by removing in that equation the convolutions with the parton density
functions, and multiplying by $x$. Furthermore, in our expressions the summation over initial-state
quark and anti-quark flavours is not contained. However, our expressions
  in the file \prog{sig\_DY\_LO\_NLO\_NNLO.m} contain the prefactors of the
quantities $x \Delta(x)$ collected in \verb|coeff[]|.  In
Tab.~\ref{tab::DY_expr} we provide translation rules from our analytic
expressions to the notation used in Eq.~(A.20) of Ref.~\cite{Hamberg:1990np}.

\begin{table}[t]
      \centering
    \begin{tabular}{c|c|c|l}
      \hline
      channel & \prog{sig\_DY\_LO\_NLO\_NNLO.m} & lines of (A.20) in~\cite{Hamberg:1990np} & contains \\
      \hline
    $\tilde{\sigma}^{(0)}_{q\bar{q}}$ & \verb|rsigc["LO", "qb"]| & 2 &
    $\Delta^{(0)}_{q\bar{q}} = \delta(1-x)$ \\
    $\tilde{\sigma}^{(1)}_{q\bar{q}}$ & \verb|rsigc["NLO", "qb"]| & 2 & $\Delta^{(1)}_{q\bar{q}}$ \\
    $\tilde{\sigma}^{(1)}_{qg}$ & \verb|rsigc["NLO", "qg"]| & 6 & $\Delta^{(1)}_{qg}$ \\
    $\tilde{\sigma}^{(2)}_{gg}$ & \verb|rsigc["NNLO", "gg"]| & 11 & $\Delta^{(2)}_{gg}$ \\
    $\tilde{\sigma}^{(2)}_{q\bar{q}}$ & \verb|rsigc["NNLO", "qb"]| & 2-5 & $\Delta^{(2)}_{q\bar{q}},\Delta^{(2)}_{q\bar{q},B^2},\Delta^{(2)}_{q\bar{q},BC},\Delta^{(2),\rm V}_{q\bar{q},AB}$ \\
    $\tilde{\sigma}^{(2)}_{qg}$ & \verb|rsigc["NNLO", "qg"]| & 6 & $\Delta^{(2)}_{qg}$ \\
    $\tilde{\sigma}^{(2)}_{qq^\prime}$ & \verb|rsigc["NNLO", "qp"]| & 7,8 & $\Delta^{(2)}_{qq,C^2},\Delta^{(2),\rm V}_{q\bar{q},CD}$ \\
    $\tilde{\sigma}^{(2)}_{qq}$ & \verb|rsigc["NNLO", "qq"]| & 9,10 & $\Delta^{(2)}_{qq,CE},\Delta^{(2)}_{qq,CF}$ \\
    \hline
  \end{tabular}
  \caption{\label{tab::DY_expr}
    The second column contains the notation for the order and channel shown in the first
    column as used in the file \prog{sig\_DY\_LO\_NLO\_NNLO.m}.
    The third column indicates in which lines of Eq.~(A.20) from
    Ref.~\cite{Hamberg:1990np}
    one can find the corresponding expression. The last column
    lists the expressions of Ref.~\cite{Hamberg:1990np} one has to consider in
    order to get our result for the partonic cross section of the second
    column.
  }
\end{table}

We are now in the position to evaluate all convolutions needed up to
N$^3$LO using \prog{MT}. As a first example we consider
the convolution
\begin{eqnarray}
  \frac{\tilde{\sigma}^{(1)}_{q\bar{q}}}{x} \otimes P_{\rm ns}^{(1)}
  \,,
\end{eqnarray}
which is a contribution to NNLO. Let us emphasize once more that
$\tilde{\sigma}^{(1)}_{q\bar{q}}$ is proportional to
$x\Delta_{q\bar{q}}^{(1)}$. The corresponding \prog{Mathematica} session looks
as follows:
\begin{mmacode}\setcounter{mmano}{0}
\In
\begin{minipage}[t]{\codewidth}%
\begin{verbatim}
<<MT`
\end{verbatim}
\end{minipage}
\\
\In
\begin{minipage}[t]{\codewidth}%
\begin{verbatim}
<<sig_DY_LO_NLO_NNLO.m
\end{verbatim}
\end{minipage}
\\
\In
\begin{minipage}[t]{\codewidth}%
\begin{verbatim}
<<splitfnsMVV.m
\end{verbatim}
\end{minipage}
\\
\In
\begin{minipage}[t]{\codewidth}%
\begin{verbatim}
Convolution[(rsigc["NLO", "qb"]/x)/.{CF->4/3, CA->3, TF->1/2},
            Pqq1/.splitfnsMVV, x];
\end{verbatim}
\end{minipage}
\\
\In
\begin{minipage}[t]{\codewidth}%
\begin{verbatim}
Collect[%, {coeff[_], ep, Lv, HPL[__], PlusDistribution[__],nl}, Together]
\end{verbatim}
\end{minipage}
\\
\Out 
$ \displaystyle {\rm
coeff("Cii[qi,bj]*(vi{}^\wedge{}2+ai{}^\wedge{}2)*alphas/Pi")} \bigg(\newline
\ldots-\frac{32 (2 x^2+1) \H_{2}(x)}{9 (x-1)}
+\ldots +{\rm Lv} \bigg(\ldots-\frac{32}{9} (x+1) \H_{1}(x)+\frac{2}{27} (8
\pi^2-27) \delta(1-x)+\ldots\bigg)+\ldots\newline
+{\rm ep}\bigg(\ldots-\frac{256}{9} (x+1) \H_{1,1,1}(x)+\ldots+\frac{4}{27}
(-96
\zeta(3)-48+19 \pi^2) \Big[\frac{1}{1-x}\Big]_+\newline
-\frac{2 (-72 x^2 \zeta(3)+\ldots+60)}{27 (x-1)} \bigg)\newline
+{\rm ep}^2\bigg( \ldots-\frac{320 (x^2+1) \H_{1,2,1}(x)}{9
  (x-1)}+\ldots\newline
+{\rm Lv}^3 \bigg(\ldots-\frac{16}{27} (x+1) \H_{1}(x)+\frac{1}{81} (8
\pi^2-27)
\delta(1-x)+\ldots\bigg)+\ldots
+\frac{160}{27} \Big[\frac{\log^4(1-x)}{(1-x)}\Big]_+\newline+\ldots
\bigg)
\bigg) 
$
\end{mmacode}
As one can see, all prefactors are collected in the arguments of the function
$\verb|coeff[]|$.
Furthermore, the final result is
available up to order $\epsilon^2$ which is needed since the convolution
with $P_{\rm ns}^{(1)}$ is
accompanied by a pole in $\epsilon$ and the NNLO expression gets
multiplied by a renormalization constant which also has an $1/\epsilon$-term.

One of the most involved convolutions which we have to consider is
\begin{eqnarray}
  \frac{\tilde{\sigma}^{(2)}_{q\bar{q}}}{x} \otimes P_{\rm ns}^{(1)}
  \,,
\end{eqnarray}
since $\sigma^{(2)}_{q\bar{q}}$ contains HPLs up to order four and plus
distributions up to $\left[ \frac{\ln^4(1-x)}{1-x}\right]_+$.
Using \prog{MT} the calculation looks as follows
\begin{mmacode}
\In
\begin{minipage}[t]{\codewidth}%
\begin{verbatim}
Convolution[(rsigc["NNLO","qb"]/x)/.{CF->4/3,CA->3,TF->1/2},
            Pqq1/.splitfnsMVV,x];
\end{verbatim}
\end{minipage}
\\
\In
\begin{minipage}[t]{\codewidth}%
\begin{verbatim}
Collect[%,{coeff[_],ep,Lv,HPL[__],PlusDistribution[__],nl},Together]
\end{verbatim}
\end{minipage}
\\
\Out 
$ \displaystyle
\rm{coeff("delta[i,j]*Sum[k,l,Q]*Cff[qk,bl]*(vk{}^\wedge{}2+ak{}^\wedge{}2)*alphas
{}^\wedge{}2/Pi{}^\wedge{}2")}
(\ldots)\phantom{\bigg(\bigg)}\newline
+\rm{coeff("delta[i,j]*Sum[k,QB]*(Cif[qi,bk]+Cif[bi,qk])*(vi{}^\wedge{}2+ai{}^\wedge{}2)}\newline{\rm *alphas{}^\wedge{}2/Pi{}^\wedge{}2")}
(\ldots)\phantom{\bigg(\bigg)}\newline
+\rm{coeff("Cii[qi,bj]*(vi{}^\wedge{}2+ai{}^\wedge{}2)*alphas{}^\wedge{}2/Pi{}^\wedge{}2")}
\bigg(\newline \ldots+\bigg(\frac{8 {\rm nl} (11 x^2+12)}{27 (x-1)}-\frac{4
  (605
x^2-922 x+617)}{27 (x-1)}\bigg) \H_{1,2}(x)\newline
+\ldots+Lv^2 \bigg(\ldots-\frac{2}{27} nl (x+5)+\ldots\bigg)+\ldots\newline
+{\rm ep}\bigg( \ldots -\frac{4 (593 x^2+635) \H_{1,0,0,0,0}(x)}{27
  (x-1)}-\frac{4
(1411 x^2+1507) \H_{1,1,0,0,0}(x)}{27 (x-1)}+\ldots\newline
+\bigg(\frac{464 {\rm nl}}{243}+\frac{8}{243} (1712 \pi^2-429)\bigg)
\Big[\frac{\log^3(1-x)}{(1-x)}\Big]_+
+\bigg(\frac{40}{27}-\frac{80 {\rm nl}}{27}\bigg)
\Big[\frac{\log^4(1-x)}{(1-x)}\Big]_+ \newline
-\frac{1024}{27} \Big[\frac{\log^5(1-x)}{(1-x)}\Big]_++\ldots
\bigg) \bigg) 
$
\end{mmacode}
where the notation for the symbols is given in the Appendix.


\section{\label{sec::sum}Summary}

  The main purpose of this paper is the presentation of the
  \prog{Mathematica} package \prog{MT} which can be used to compute
  convolution integrals.  An algorithm has been implemented based on Mellin
  transformation and the introduction of generalized derivatives which allows
  for a simultaneous treatment of HPLs and delta and plus distributions on
  the same footing.  \prog{MT} contains several functions to perform
  (inverse) Mellin transforms and manipulations of harmonic sums and plus
  distributions. Furthermore, Mellin transformations and convolutions can
    also be performed numerically.

To exemplify the functionality of \prog{MT} we have considered all convolution
integrals to the Higgs boson production and Drell-Yan process up to N$^3$LO in
QCD perturbation theory. As a by-product the NNLO Drell-Yan cross section has
been computed including ${\cal O}(\epsilon)$ terms.

\prog{MT} can be downloaded from the website~\cite{progdata} which also
contains the splitting functions and the partonic cross sections for the 
Drell-Yan process in computer-readable form.
The partonic cross sections for Higgs production can be found on
the website~\cite{progdata_higgs}.


\section*{Acknowledgments}

This work is supported by the Deutsche
Forschungsgemeinschaft in the Sonderforschungsbereich/Transregio~9
``Computational Particle Physics''.


\section*{\label{sec::app}Appendix: Description of \prog{sig\_DY\_LO\_NLO\_NNLO.m}}

The {\rm Mathematica} file \prog{sig\_DY\_LO\_NLO\_NNLO.m}
contains the partonic cross sections for Drell-Yan production
at LO, NLO and NNLO encoded in the functions listed in
Tab.~\ref{tab::DY_expr}. The meaning of the symbols used
in \prog{sig\_DY\_LO\_NLO\_NNLO.m} is explained in
Tabs.~\ref{tab::sig.m} and~\ref{tab::coeff} where
$C_F=4/3$, $C_A=3$ and $T_F=1/2$ are QCD color factors, $n_l$ counts the
number of massless quarks and $\epsilon=(4-D)/2$ is the regularization
parameter with $D$ being the space-time dimension.
$\mu$ is the renormalization scale and $\sqrt{Q^2}$ is the invariant mass of
the produced lepton pair.

\begin{table}[t]
  \centering
  \begin{tabular}{c|c}
    \hline
    $\verb|ep|$ & $\epsilon$ \\
    $\verb|nl|$ & $n_l$ \\
    $\verb|CF|$ & $C_F$ \\
    $\verb|CA|$ & $C_A$ \\
    $\verb|TF|$ & $T_F$ \\
    $\verb|Lv|$ & $\ln(\mu^2/Q^2)$ \\
    $\verb|coeff[]|$ &
    Labels in the argument are described in Tab.~\ref{tab::coeff}. \\
    \hline
  \end{tabular}
  \caption{\label{tab::sig.m} Notation used in \prog{sig\_DY\_LO\_NLO\_NNLO.m}.}
\end{table}

\begin{table}[t]
  \centering
  \begin{tabular}{c|c}
    \hline
    $\verb|alphas|$ & $\alpha_s$ \\
    $\verb|Sum[k,QB]|$ & $\sum_{k \in Q, \bar{Q}}$ \\
    $\verb|qi|$ & $q_i$ \\
    $\verb|bi|$ & $\bar{q}_i$ \\
    $\verb|Cii[qi,bj]|$ & coupling matrix $C^{\rm{ii}}(q_i,\bar{q}_j)$ to vector bosons\\
    $\verb|Cif[qi,qj]|$ & coupling matrix $C^{\rm{if}}(q_i,q_j)$ to vector bosons\\
    $\verb|Cff[qk,bl]|$ & coupling matrix $C^{\rm{ff}}(q_k,\bar{q}_l)$ to vector bosons\\
    $\verb|vi|$ & $v_i$\\
    $\verb|ai|$ & $a_i$\\
    $\verb|delta[i,j]|$ & $\delta_{ij}$\\
    \hline
  \end{tabular}
  \caption{\label{tab::coeff} Symbolic notation used in the argument of
    \texttt{coeff[]} in \prog{sig\_DY\_LO\_NLO\_NNLO.m}. For further
    explanations see
    also Appendix~A of Ref.~\cite{Hamberg:1990np}.}
\end{table}





\begin{thebibliography}{99}


%
%

\bibitem{Ellis:1991qj}
  R.~K.~Ellis, W.~J.~Stirling and B.~R.~Webber,
  Camb.\ Monogr.\ Part.\ Phys.\ Nucl.\ Phys.\ Cosmol.\  {\bf 8} (1996) 1.

\bibitem{Blumlein:1998if}
  J.~Blumlein and S.~Kurth,
  Phys.\ Rev.\  D {\bf 60} (1999) 014018
  [arXiv:hep-ph/9810241].

\bibitem{Ablinger:2011te}
  J.~Ablinger, J.~Blumlein and C.~Schneider,
  J.\ Math.\ Phys.\  {\bf 52} (2011) 102301
  [arXiv:1105.6063 [math-ph]].

\bibitem{Hoschele:2012xc}
  M.~H\"oschele, J.~Hoff, A.~Pak, M.~Steinhauser and T.~Ueda,
  Phys.\ Lett.\ B {\bf 721} (2013) 244
  [arXiv:1211.6559 [hep-ph]].

\bibitem{Buehler:2013fha}
  S.~Buehler and A.~Lazopoulos,
  arXiv:1306.2223 [hep-ph].

\bibitem{Altarelli:1977zs}
  G.~Altarelli and G.~Parisi,
  Nucl.\ Phys.\ B {\bf 126} (1977) 298.

\bibitem{Curci:1980uw}
  G.~Curci, W.~Furmanski and R.~Petronzio,
  Nucl.\ Phys.\ B {\bf 175} (1980) 27.

\bibitem{Moch:2004pa}
  S.~Moch, J.~A.~M.~Vermaseren and A.~Vogt,
  Nucl.\ Phys.\ B {\bf 688} (2004) 101
  [hep-ph/0403192].

\bibitem{Vogt:2004mw}
  A.~Vogt, S.~Moch and J.~A.~M.~Vermaseren,
  Nucl.\ Phys.\ B {\bf 691} (2004) 129
  [hep-ph/0404111].

\bibitem{Pak:2011hs}
  A.~Pak, M.~Rogal and M.~Steinhauser,
  JHEP {\bf 1109} (2011) 088
  [arXiv:1107.3391 [hep-ph]].

\bibitem{Vermaseren:1998uu}
  J.~A.~M.~Vermaseren,
  Int.\ J.\ Mod.\ Phys.\ A {\bf 14} (1999) 2037
  [hep-ph/9806280].

\bibitem{Remiddi:1999ew}
  E.~Remiddi and J.~A.~M.~Vermaseren,
  Int.\ J.\ Mod.\ Phys.\ A {\bf 15} (2000) 725
  [hep-ph/9905237].

\bibitem{harmpol}
\url{http://www.nikhef.nl/~form/maindir/packages/harmpol/}.

\bibitem{Maitre:2005uu}
  D.~Maitre,
  Comput.\ Phys.\ Commun.\  {\bf 174} (2006) 222
  [hep-ph/0507152].

\bibitem{Maitre:2007kp}
  D.~Maitre,
  Comput.\ Phys.\ Commun.\  {\bf 183} (2012) 846
  [hep-ph/0703052].

\bibitem{Anastasiou:2005qj}
  C.~Anastasiou, K.~Melnikov and F.~Petriello,
  Nucl.\ Phys.\ B {\bf 724} (2005) 197
  [hep-ph/0501130].

\bibitem{Dittmaier:2011ti}
  S.~Dittmaier, C.~Mariotti, G.~Passarino, R.~Tanaka {\it et al.},
   [LHC Higgs Cross Section Working Group Collaboration],
  arXiv:1101.0593 [hep-ph].

\bibitem{Dittmaier:2012vm}
  S.~Dittmaier, C.~Mariotti, G.~Passarino, R.~Tanaka {\it et al.},
   [LHC Higgs Cross Section Working Group Collaboration],
  arXiv:1201.3084 [hep-ph].

\bibitem{progdata_higgs}
{\tt http://www.ttp.kit.edu/Progdata/ttp12/ttp12-45}

\bibitem{Anastasiou:2002yz}
  C.~Anastasiou and K.~Melnikov,
  Nucl.\ Phys.\  B {\bf 646} (2002) 220,
  arXiv:hep-ph/0207004.

\bibitem{Pak:2009dg}
  A.~Pak, M.~Rogal and M.~Steinhauser,
  JHEP {\bf 1002} (2010) 025
  [arXiv:0911.4662 [hep-ph]].

\bibitem{Kramer:1986sg}
  G.~Kramer and B.~Lampe,
  Z.\ Phys.\  C {\bf 34} (1987) 497
  [Erratum-ibid.\  C {\bf 42} (1989) 504].

\bibitem{Matsuura:1987wt}
  T.~Matsuura and W.~L.~van Neerven,
  Z.\ Phys.\  C {\bf 38} (1988) 623.

\bibitem{Matsuura:1988sm}
  T.~Matsuura, S.~C.~van der Marck and W.~L.~van Neerven,
  Nucl.\ Phys.\  B {\bf 319} (1989) 570.

\bibitem{Gehrmann:2005pd}
  T.~Gehrmann, T.~Huber and D.~Maitre,
  Phys.\ Lett.\  B {\bf 622} (2005) 295
  [arXiv:hep-ph/0507061].

\bibitem{Baikov:2009bg}
  P.~A.~Baikov, K.~G.~Chetyrkin, A.~V.~Smirnov, V.~A.~Smirnov and
  M.~Steinhauser,
  Phys.\ Rev.\ Lett.\  {\bf 102} (2009) 212002
  [arXiv:0902.3519 [hep-ph]].

\bibitem{Gehrmann:2010ue}
  T.~Gehrmann, E.~W.~N.~Glover, T.~Huber, N.~Ikizlerli and C.~Studerus,
  JHEP {\bf 1006} (2010) 094
  [arXiv:1004.3653 [hep-ph]].

\bibitem{Lee:2010cga}
  R.~N.~Lee, A.~V.~Smirnov and V.~A.~Smirnov,
  JHEP {\bf 1004} (2010) 020
  [arXiv:1001.2887 [hep-ph]].

\bibitem{Laporta:2001dd}
  S.~Laporta,
  Int.\ J.\ Mod.\ Phys.\ A {\bf 15} (2000) 5087
  [hep-ph/0102033].

\bibitem{Anastasiou:2012kq}
  C.~Anastasiou, S.~Buehler, C.~Duhr and F.~Herzog,
  JHEP {\bf 1211} (2012) 062
  [arXiv:1208.3130 [hep-ph]].

\bibitem{Hamberg:1990np}
  R.~Hamberg, W.~L.~van Neerven and T.~Matsuura,
  Nucl.\ Phys.\ B {\bf 359} (1991) 343
   [Erratum-ibid.\ B {\bf 644} (2002) 403].

\bibitem{Harlander:2002wh}
  R.~V.~Harlander and W.~B.~Kilgore,
  Phys.\ Rev.\ Lett.\  {\bf 88} (2002) 201801,
  arXiv:hep-ph/0201206.

\bibitem{progdata}
{\tt http://www-ttp.kit.edu/Progdata/ttp13/ttp13-27}


\end{thebibliography}
\end{document}